# 8×8 Reconfigurable quantum photonic processor based on silicon nitride waveguides

**Caterina Taballione,** [1,2*] **Tom A. W. Wolterink,** [3] **Jasleen Lugani,** [3] **Andreas Eckstein,**[3] **Bryn A. Bell,**[3] **Robert Grootjans,**[4] **Ilka Visscher,**[4] **Dimitri Geskus,**[4] **Chris G. H. Roeloffzen,**[4] **Jelmer J. Renema,** [2,3,5] **Ian A. Walmsley**[3]**, Pepijn W. H. Pinkse**[5]**, and Klaus-J. Boller**[1]

[1]*Laser Physics and Nonlinear Optics (LPNO), University of Twente, PO Box 217, 7500 AE Enschede, The Netherlands*
[2]*QuiX BV, PO Box 8505, Veilingstraat 7, 7545 LZ Enschede, The Netherlands*
[3]*Department of Physics, University of Oxford, Clarendon Laboratory, Parks Road, Oxford, OX1 3PU, United Kingdom*
[4]*Lionix International BV, PO Box 456, 7500 AL Enschede, The Netherlands*
[5]*Complex Photonic Systems (COPS), University of Twente, PO Box 217, 7500 AE Enschede, The Netherlands*
*\*c.taballione@utwente.nl*

**Abstract:** The development of large-scale optical quantum information processing circuits ground on the stability and reconfigurability enabled by integrated photonics. We demonstrate a reconfigurable 8×8 integrated linear optical network based on silicon nitride waveguides for quantum information processing. Our processor implements a novel optical architecture enabling any arbitrary linear transformation and constitutes the largest programmable circuit reported so far on this platform. We validate a variety of photonic quantum information processing primitives, in the form of Hong-Ou-Mandel interference, bosonic coalescence/anti-coalescence and high-dimensional single-photon quantum gates. We achieve fidelities that clearly demonstrate the promising future for large-scale photonic quantum information processing using low-loss silicon nitride.

## 1. Introduction

Optical quantum information processing (QIP) aims to solve computational tasks such as quantum simulation[1-4] and quantum machine learning [5, 6] more efficiently than can be done with classical computers, using photons as information carriers and large-scale interferometers as processors. To perform increasingly complex tasks on increasingly complex interferometers, maintaining optical path-length stability on a large scale is a central technological challenge.

The key enabling technology for optical QIP is integrated photonics, as it allows linear optical networks to have interferometric stability and, in addition, reconfigurable functionality, via tunable beam splitters and phase shifters. Integrated photonics permit, moreover, the realization of other key components such as squeezed light sources [7], homodyne detection systems [8] and fast switching [9] that provide a large benefit to quantum information based on continuous variables [10-13].

On-chip linear optical networks enable a variety of quantum information and communication protocols such as quantum teleportation [14], on-chip quantum key distribution [15, 16], photonic quantum gates [17] and boson sampling [18-21]. Such structures are also exploited for classical applications, such as machine learning [5, 22-24] and signal processing in microwave photonics [25]. Imperative requirements for on-chip linear optical networks are low propagation loss and high component density.

Amongst the silicon-based platforms, stoichiometric silicon nitride ($Si_3N_4$) offers the unique combination of high index contrast [26] for low bend loss and ultra-low straight-propagation loss [27, 28], enabling thus both dense and low-loss linear optical networks on a compact footprint (see Fig. 1). $Si_3N_4$ is a well-known platform for many classical applications [25, 29-32] and shows a mature state-of-the-art technology [33]. Increasingly, silicon nitride is attracting interest in the field of quantum optics for its tremendous potential in implementing on-chip single-photon sources [34, 35], squeezed light [36-39], superconductive single-photon detectors [40, 41], quantum walks [42] and time-bin encoding for quantum communication [43, 44]. Moreover, due to its wide spectral transparency range (from 440 nm to 2.5 μm), silicon nitride allows interfacing with all common quantum light sources, e.g., spontaneous parametric down-conversion, spontaneous four-wave mixing sources and quantum dots.

A clear demonstration of the suitability of this promising platform for realizing large-scale universal programmable linear optical networks is, however, still missing.

In fact, doped silica and silicon-on-insulator technologies host, so far, the majority of on-chip universal linear optical networks for QIP [45]. While the doped silica platform offers low propagation loss only with a larger footprint, i.e., low component density, silicon-on-insulator allows for dense optical circuits at the expense of higher loss. As large scale photonic circuits for QIP require both low propagation losses and high component density, a platform that combines these properties is very beneficial for the development of QIP. This combination is what silicon nitride offers.

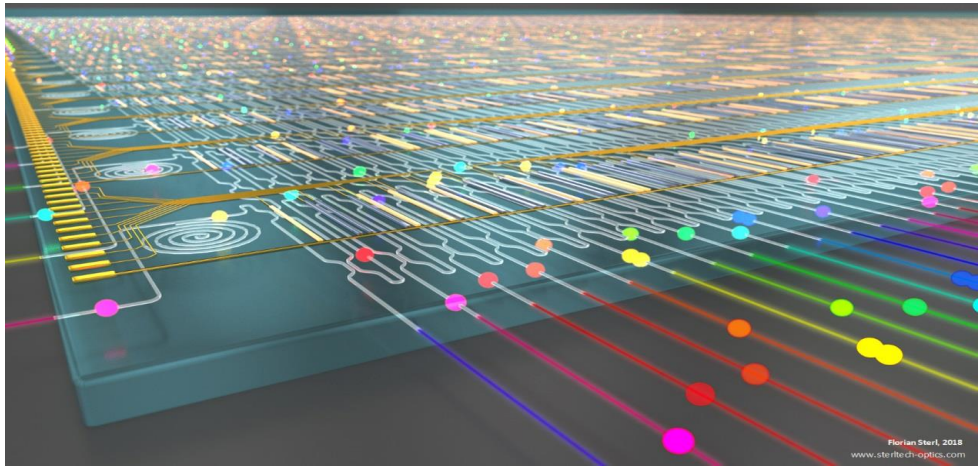

Fig. 1. Schematic geometrical layout of the programmable silicon nitride linear optical waveguide network where tight guiding in high-contrast SiN waveguides is exploited to form a dense waveguide mesh, with currently 21 Mach-Zehnder interferometers per square centimeter. Bullets indicate the injection of single photons via the front or side facets of the chip. The circuit contains delay lines to equalize photon arrival and departure times at the front and back ports of the internal interferometers. The location of electric heaters with feed wires and bond pads is shown at the top surface in gold. Choosing bullets with different colors indicates that the wide transparency range of the waveguides allows using the processor with quantum light sources over a wide spectral range, from the visible to the mid-infrared.

Here, we demonstrate, for the first time, an 8×8-mode linear optical network implemented in a fully reconfigurable photonic processor based on silicon nitride waveguides. The photonic processor contains the highest density of components to date compared to other state-of-the-art processors in other platforms [1, 17, 46], considering the number of optical components that can be placed sequentially until the power transmission through the processor is reduced to a targeted arbitrary value $e^{-1}$ (for further discussion see section 5.1 in Appendix). The lowest loss measured so far on the silicon nitride platform, in almost straight sections, is 0.045 dB/m [27], while the lowest loss at sharp curvature has been shown to be as low as 0.13 dB/m at 115 μm radius of curvature [28]. In the Appendix we present an overview on previously measured loss data vs curvature, which enables to estimate the lowest loss in circuits once the design of a circuit is specified (see section 5.4 in Appendix). The 8×8-mode photonic processor, the largest realized so far in silicon nitride, includes 128 reconfigurable elements: 64 tunable beam splitters, constructed as Mach-Zehnder interferometers with internal thermo-optic phase shifters, and 64 additional phase shifters, arranged in a novel linear optical architecture. The processor contains the optical realization of a Blass matrix [47, 48], a well-known architecture for beamforming networks in microwave engineering, where it is used for directional transmission of radio frequency signals to and from antenna arrays. Translating this architecture from microwave engineering to optical QIP, the Blass matrix supports the realization of any arbitrary linear transformation, both unitary [49, 50] and non-unitary [51-53]. We show that our processor preserves the coherence of quantum states by programming the processor to implement quantum interference. As a proof-of-principle demonstration of the architecture's capability to implement non-unitary transformations, we show anti-coalescence of bosons [54-56] on a 2×2 Blass matrix. Finally, we realize high-dimensional single-photon quantum gates exploiting the whole spatial mode structure of the processor.

## 2. Experimental setup

Figure 2 shows a schematic of the experimental setup. The photonic processor (Fig. 2(a)) consists of 64 unit cells, each composed of a phase shifter (in red) and a tunable beam splitter (in blue), in an arrangement that enables any arbitrary 8×8 transformation. Each tunable beam splitter is constructed as a Mach-Zehnder interferometer, based on two directional couplers and an internal thermo-optic phase shifter. The 128 thermally tunable elements are remotely controllable and are designed to each allow a $\frac{3}{2}\pi$ phase shift. The tuning range can be extended with longer tunable elements or stronger current supplies.

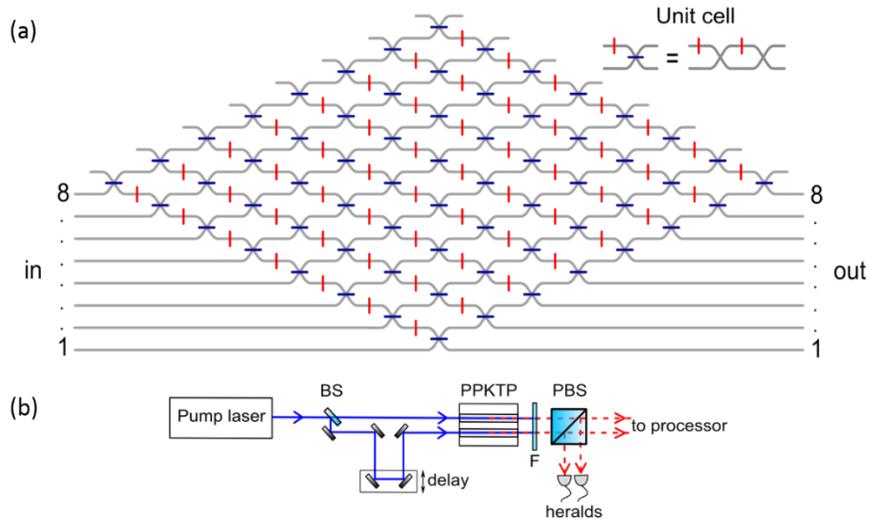

Fig. 2. Schematic of the experimental setup. (a) The photonic processor is composed of 64 unit cells, each comprising a phase shifter (red vertical line) and a tunable beam splitter (blue horizontal line) implemented as a Mach-Zehnder interferometer. The rhomboidal shape of the processor's schematic has been chosen for better overview. In the real processor the elements are arranged on a square mesh. (b) Photon pairs are generated via type-II parametric down-conversion in PPKTP waveguides pumped with a mode-locked laser at 775 nm and injected into the photonic processor.

The photonic processor is based on stoichiometric silicon nitride waveguides, grown with low-pressure chemical vapor deposition, with a double-stripe cross-section [33]. The waveguides exhibit a propagation loss of 0.2 dB/cm for a total on-chip transmission greater than or equal to 60%, a value that corresponds to the longest optical path. The coupling losses to a single-mode optical fiber are about 2.9 dB/facet which can be greatly reduced (to 0.5 dB/facet [33]) by waveguide tapering.

Single photons for the experiments are provided with two parametric down-conversion sources (Fig. 2(b)). Frequency-doubled light from a mode-locked fiber laser, with a center wavelength of 775 nm and a spectral width of 2 nm, is divided into two paths, one containing an adjustable delay line, and focused into two 10-mm-long periodically-poled $KTiOPO_4$ (PPKTP) waveguides [57]. Each PPKTP waveguide generates, via type-II down-conversion, orthogonally-polarized spectrally-separable photon pairs at telecom wavelengths (signal and idler at 1547 nm and 1553 nm, respectively). After removal of the pump wavelength (F filter for pump rejection) the signal and idler photons are separated using a polarizing beam splitter (PBS) and collected by single-mode fibers. The signal photons are used to herald the idler photons with a heralding efficiency of 30%. The two idler photons are coupled into the photonic processor using polarization-maintaining fibers. For photodetection, we use a set of fiber-coupled superconducting single-photon detectors (efficiency 85% [58]).

## 3. Results

### 3.1 Characterization of tunable elements

In order to prepare the processor to perform a desired task, we have to characterize the response of each tunable elements. The calibration of the tunable elements requires to measure the phase shift induced by the applied heating voltage, U. This voltage dependence follows a square law, $\phi(U) = c + d \cdot U^2$, which was measured for each phase shifter by reading interference fringes in the bar mode photon count value of the respective MZI (see inset Fig. 3 for bar and cross definition). The theoretically expected response at the bar mode is a sinusoidal function, $f = a - b \cdot \cos(\phi(U))$. The calibration parameters a, b, c and d are real numbers and are determined by a least-squares fit. Each tunable element is calibrated independently, as described in the supplement of [17], starting, in our case, from the tunable beam splitter at the bottom corner of the processor through which light can be directed without having to pass other elements.

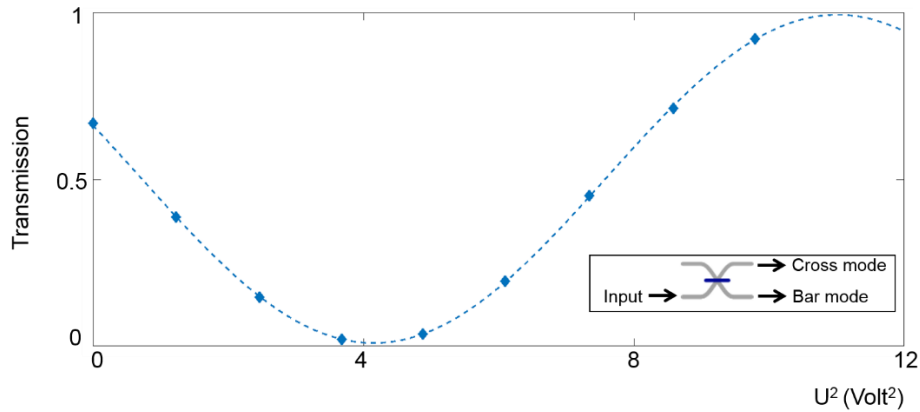

Fig. 3. Transmission of the first accessible tunable beam splitter. The diamonds are the experimental data and the blue dashed line is the sinusoidal fit. The visibility of the sinusoid gives the range of splitting ratio achievable and the period of the sinusoid is related to the phase shift induced by the tunable element.

Figure 3 shows, as an example, the transmission at the bar mode of the first accessible beam splitter (bottom corner in Fig. 3) versus the square of the applied heating voltage, as compared to the theoretically expected response f, which is fit to the data. The residual deviations of the data from the sinusoidal curve indicates that the beam splitter can be tuned very precisely to any desired phase value, within its tuning range. As can be seen, a full period of sinusoidal curve is not achievable. This limitation can be removed both by making longer heaters and by allowing for higher currents. The distribution of the splitting ratios of the 128 directional couplers can be described by an average value of 0.497 with a standard deviation of 0.126. The average phase range is ≈4.59 rad (~ $\frac{3\pi}{2}$) and the average splitting ratio of the tunable beam splitters, i.e., MZI, is adjustable between 5% and 95%. The distribution of the phase offset at the MZIs has a mean value of ≈3.5 rad with a standard deviation of 1.85 rad. Taking into account the limited phase range and the phase offset, we find that the tunable beam splitters of the processor, i.e., the MZI, can always achieve a splitting ratio of 50:50 either for an internal phase shift $\phi = \frac{\pi}{2}$ or $\phi = \frac{3\pi}{2}$.

## 3.2 On-chip quantum interference

To demonstrate the suitability of the photonic processor for QIP, we first observe Hong-Ou-Mandel (HOM) interference [59] between two photons (bullets of the same color in Fig. 4(a)) at various positions (beam splitters) within the processor (colored disks in Fig. 4(a)).

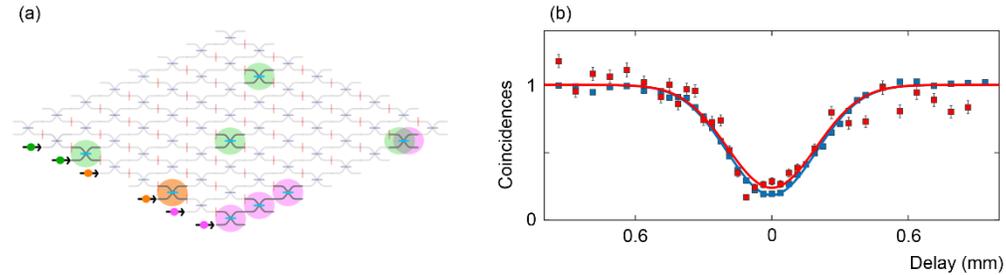

Fig. 4. (a) Two-photon interference at various locations of the processor (colored circles), also indicating the used pairs of input waveguides (bullets of the same color). (b) Coincidence probability versus delay. The two-photon interference measured at the tunable beam splitters on the processor (red data points) is well in accord with the off-chip reference measurement (blue data points). All the investigated beam splitters show a similar visibility. The solid curves indicates Gaussian fits to the data. The error bars are given by the square root of the number of coincidences.

The processor is configured to route the two incident photons across the chip to a targeted beam splitter, which is programmed to a reflectivity of 50%, after which the photons are directed to two outputs. The coincidence count rates at the outputs versus the relative delay of the two single photons are recorded. In Fig. 4(b) we show, as an example, the HOM interference at one of the targeted tunable beam splitters (red curve), at the center of the processor in comparison with a reference measurement, i.e., an off-chip HOM experiment using a fiber beam splitter (blue curve). It can be seen that the two measurements are well in accord. At a mean photon number of 0.01 we measure a reference HOM dip visibility $V_{\text{ref}}$ of 81% between the idler photons of the two sources, with the visibility defined as $V = \left|\frac{C_d - C_i}{C_d}\right|$, where $C_d$ and $C_i$ are respectively the coincidence counts for temporally distinguishable and indistinguishable photons. We repeat the experiment at various positions, i.e., beam splitters, within the photonic processor obtaining similar results, i.e., an average visibility of 76%. The consistency of the measured on-chip HOM dips, over the whole processor depth, with the reference shows that our photonic processor preserves the spectro-temporal similarity of the photons and confirms the suitability of the photonic processor for quantum information processing. The difference of visibility between the on- and off-chip case is probably to be addressed to the residual on-chip path length difference between the optical path from each input port to the targeted beam splitter.

*3.3 Arbitrary linear transformations*

Due to its architecture, the photonic processor can be configured to perform arbitrary linear transformations on its 8 modes, both unitary and non-unitary [51-53], the latter implemented via ancillary modes. In QIP, non-unitary, lossy, transformations are typically considered detrimental. However, the additional freedom obtained by removing the restriction of unitarity allows for new transformations that exhibit exciting behavior such as a tunable quantum interference and an apparent nonlinear absorption [55, 60, 61]. Already the simple case of a balanced symmetric lossy beam splitter contains free parameters determining the relative phase of the transmission coefficients that enable the tuning of the well-known HOM-like dip, the signature of bosonic coalescence, into a HOM-peak for bosonic anti-coalescence.

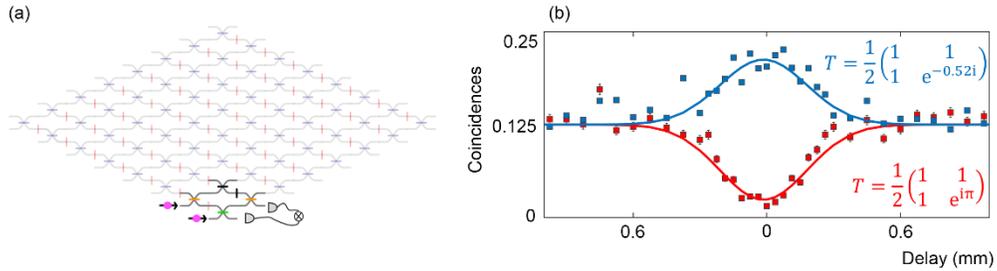

Fig. 5. (a) Implementation of a lossy beam splitter on a 2×2 Blass matrix. The black elements are set accordingly to the desired T. (b) Coincidence versus delay for two different lossy transformations T. Measured two-photon bosonic coalescence (red curve)/anti-coalescence (blue curve) with a visibility of 81% and 70% respectively.

To illustrate how the Blass matrix architecture allows the implementation of non-unitary transformations, we realize a balanced symmetric lossy beam splitter, involving four beam splitters and two phase shifters (see Fig. 5(a)), described by the matrix $T = \frac{1}{2}\begin{pmatrix} 1 & 1 \\ 1 & e^{i\alpha} \end{pmatrix}$, with the phase $\alpha$ a free parameter. The behavior observed in a quantum interference experiment between two single photons will oscillate between coalescence and anti-coalescence of the photons depending on this phase $\alpha$ [61], with the well-known HOM-like coalescence for $\alpha = \pi$. The photonic processor is programmed to perform such a non-unitary 2×2 transformation on a 2×2 Blass matrix. Fig. 5(b) shows the quantum interference between two single photons for two different non-unitary 2×2 transformations implemented on the chip, resulting in bosonic coalescence for phase $\alpha = \pi$ (red) and anti-coalescence for $\alpha = -0.52$ rad (blue). The visibility of these HOM-like dip and peak are 81% and 70%, respectively, as expected for these specific transformations.

*3.4 High-dimensional quantum logic gates*

High-dimensional quantum states, i.e., qudits, are of importance for large-alphabet quantum communication protocols [62] and cryptography [63]. In optics, qudits can be implemented using a modal degree of freedom, spatial or temporal, of the single photon to encode information. When encoding in the spatial degrees of freedom, either via transverse spatial modes [64] or rail-encoded modes, large unitary linear optical networks can be exploited to implement high-dimensional quantum logic gates for the control and manipulation of such qudits [65, 66]. As shown in [65], providing all the integer powers of a d-dimensional X-gate, i.e., $X, X^2 ... X^d$, and of a d-dimensional Z-gate, enables any unitary operation in a d-dimensional state space, with d=8 in our case, where the action of a d-dimensional X-gate can be described as $X|j\rangle = |j+1 \bmod d\rangle$.

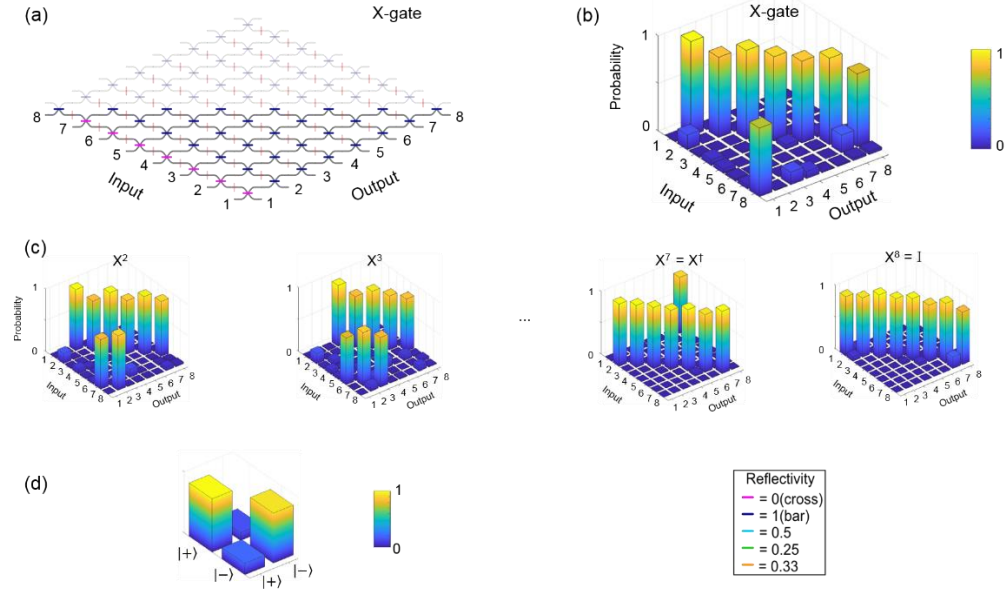

Fig. 6. (a) Realization of an 8-dimensional X-gate and (b) its measured truth table. (c) Truth tables of integer powers of the X-gate reported above. The average fidelity is $\mathcal{F}_{8\times 8} = 94.6\%$. (d) Evolution of a coherent superposition input state $|\pm\rangle = (1/\sqrt{2})(|1\rangle_5 \pm |1\rangle_6)$ through a 6-dimensional X-gate, giving a fidelity of 91.9%.

Here we demonstrate the realization of an 8-dimensional X-gate (see Fig. 6(a)) and all its integer powers, i.e., $X, X^2 \ldots X^8$, in an 8-dimensional-rail encoding thus exploiting the whole mode structure of the processor. Figure 6(b) shows the measured truth table for the X-gate, obtained by injecting single photons into each of the 8 inputs. The results for the integer powers of the X-gate are summarized in Fig. 6(c). The fidelities of these gates are about $\mathcal{F}_{8\times 8}=94.6\%$, where the fidelity of each gate is calculated as the average state fidelity $\mathcal{F} = \sum_i \frac{\sqrt{p_i^{exp} \cdot p_i^{th}}}{8}$, with $p_i^{th}$ and $p_i^{exp}$ being the theoretical and experimental probabilities for each computational input $i$, respectively. Finally, we measure the transformation of a single photon in the coherent superposition state $\frac{1}{\sqrt{2}}(|1\rangle_5 \pm |1\rangle_6)$ through a 6-dimensional X-gate, with a measured gate fidelity of $\mathcal{F}_{6\times 6}=96.2\%$. Figure 6(d) shows the action of the 6-dimensional X-gate on the coherent superposition input state showing that the gate preserves the relative phase of the state.

## 4. Conclusions

We report the realization of a fully programmable and remotely controllable 8×8-mode photonic processor, which is the largest universal linear optical network realized on $Si_3N_4$. We have demonstrated a variety of QIP primitives such as on-chip HOM interference, bosonic anti-coalescence on a 2×2 Blass matrix and high-dimensional single-photon quantum gates. The obtained results show that our processor retains the indistinguishability of the photons, limited only by the off-chip single-photon source, and enables any arbitrary linear transformation. Our findings demonstrate the promising future of the $Si_3N_4$ platform for the development of large reconfigurable universal linear optical quantum circuits.

## 5. Appendix

*5.1 Functional complexity*

In the variety of on-chip universal linear optical networks presented so far, two conflicting developments can be recognized. To achieve higher degrees of functionality, i.e., higher functional complexity, an increasing number of functional elements, e.g., tunable or switchable, is required. Since the size of integrated optical chips is constrained to a chip or wafer by fabrication technology, the density of components on photonic chips can ultimately only be increased via a reducing the size of components and making sharper bends in the waveguides. On the other hand, it remains a central requirement to maintain the lowest propagation loss also with a growing number of components and therefore growing optical path lengths, particularly for quantum processing schemes. Since tunability, component size and optical loss are intrinsically coupled properties of the optical materials and fabrication technology used, the future of photonic processors depends critically on which material platform will enable the greatest functional complexity for a given level of acceptable loss.

An impressive variety of quantum photonic processors has been demonstrated with different material platforms. A most prominent representative for semiconductor materials is silicon-on-insulator (SOI) as employed for demonstration of, e.g., bosonic transport simulation [1]. The advantage of the SOI platform is that it supports extremely dense photonic circuits, thanks to its high index contrast between the waveguide core and cladding ($\Delta n \approx 40\%$ [26]), because this allows for small feature size via tight bending radii without much radiation loss. Also, due to the relatively small bandgap, highly responsive tuning elements, with typically $\pi$ phase shift within less than 100 µm propagation length, can be realized using carrier injection [67]. On the other hand, surface scattering in combination with crystallinity and the high index contrast enhances the optical propagation loss in SOI to levels of several dB/cm, both in straight and curved waveguide sections. Additional loss occurs as the drawback of high responsivity in tuning, due to free-carrier absorption [68].

In contrast to semiconductors, amorphous dielectric materials with large bandgap are known for lowest propagation loss, the most well-known representative being doped silica. This platform has been used, for instance, for demonstrating universal linear optical circuits [17]. The typical propagation loss is at least an order of magnitude lower than in SOI, at the level of 0.1 dB/cm [26], and thermal tuning can be applied without inducing noticeable additional loss. However, the disadvantage is the low index contrast which is inherent to waveguide cores based on doping ($\Delta n \approx 0.5\%$ [26]). The low contrast leads to weak guiding and severe radiation loss occurs at smaller curvatures, which makes dense and thus complex waveguide circuits unfeasible.

The work we present here makes use of an advanced dielectric waveguide platform involving waveguide cores made from stoichiometric silicon nitride, embedded in a cladding made from stoichiometric silicon oxide. Based on slow deposition at high temperatures (low-pressure chemical vapor deposition), these waveguides offer a unique combination of high index contrast ($\Delta n \sim 18\%$ [26]) and ultralow propagation loss (> 0.0004 dB/cm [27]). The platform thus offers ideal preconditions for the realization of dense and low-loss photonics circuits, where tuning-induced loss can be neglected.

In the following, we inspect the different degrees of complexity achievable with the named photonic platforms. For a quantitative comparison, we estimate the maximum achievable complexity with SOI, doped silica and $Si_3N_4$ in terms of a new figure of merit, $C_f$. We define this figure as the maximum number of functional unit cells, $n^2$, that can be arranged in a 2D square mesh, before the intensity of the light has dropped to a fraction f. For convenience and definiteness, we proceed with a specific value for this fraction, $f = e^{-1}$; however, any other value can be selected as desired.

Next, for comparing universal photonic processors, we define a prototype waveguide circuit as unit cell, with many unit cells forming the processor. As has been shown, all unitary transformations can be implemented using a network of unit cells that can perform two essential functions, tunable beam splitting followed by tunable phase shifting [49].

The most basic functional design of such a unit cell in the form of a 2D waveguide circuit is displayed in Fig. 7. We consider the beam splitter realized with two directional couplers in the form of a Mach-Zehnder interferometer (MZI) where the splitting ratio can be tuned via a phase shifter in one of the interferometer arms. An additional phase shifter in one of the output waveguides allows tuning of the relative phase between the two outputs of the MZI. The optical loss caused by a single unit cell depends on some basic geometrical parameters, specifically, the propagation length through straight waveguide sections where phase tuning is provided, $L_t$, and the propagation length through sections that are bent with a certain radius R, $L_b = 4\pi R$.

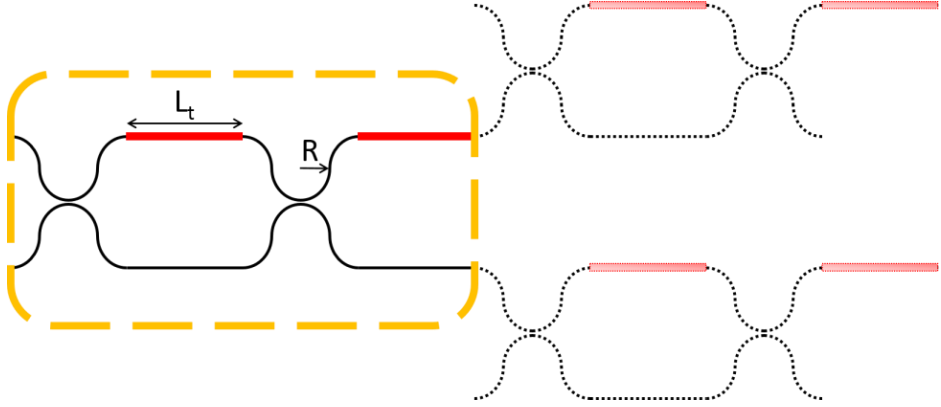

Fig. 7. Schematic of a universal linear optical network. The unit cell (dashed frame) comprises a tunable beam splitter in the form of a tunable Mach-Zehnder interferometer with two directional couplers and a phase shifter in one arm of length $L_t$ (red), followed by an external phase shifter ($L_t$, red). The path length through the unit cell is determined by the length of the straight-waveguide tunable elements and the length of curved waveguides (radius of curvature R).

The optical loss associated with the various waveguide sections depend on the chosen waveguide platform as described above and on the chosen core cross section of the waveguides. The latter dependence means that the intrinsically lowest-loss of a specified circuit fabricated with a given material platform can only be approached by a variation of the core cross section along propagation. Specifically, for each waveguide section depending on its curvature, a different cross section has to be chosen that minimizes the propagation loss. In section 5.4 we describe how we have obtained for each of the three platforms an empiric relation between the waveguide bending radius and the minimum achievable power loss constant, $\alpha_b(R)$, specified in unit of dB/m. Having defined a waveguide circuitry for the unit cell of a photonic processor in terms of waveguide lengths and curvatures then allows to estimate the number of unit cells to be passed before the specified loss fraction, f, is reached.

When analyzing the generic unit cell shown in Fig. 7, its path length, over which light propagates, is $L_{uc} = L_b + 2L_t$. Here we have assumed for simplicity that the circuit makes use of 90°-bends and that the length of the directional couplers can be neglected compared to the length of tunable and other straight sections.

The number of unit cells, n, that can be coupled in series until the power transmission is reduced to $e^{-1}$ is found by solving the following equation $T_n = \exp[-(n \cdot \alpha \cdot L \cdot \ln(10))/10] = e^{-1}$ for n. In this expression α specifies the power loss for the sections with straight-propagation, with phase-tuning and for bent waveguide sections, respectively $\alpha_s$, $\alpha_t$ and $\alpha_b$, and where $L_s$, $L_t$, and $L_b$ are the according lengths of the sections. The total propagation losses are thus given by $\alpha \cdot L = \alpha_s \cdot L_{uc} + \alpha_b \cdot L_b + \alpha_t \cdot (2L_t)$. The number of unit cells that can be arranged in a 2D square mesh obeying the same loss condition, i.e., the figure of merit for maximum functional complexity, becomes $C_f = n^2 = (10/(\alpha \cdot L_{uc} \cdot \ln(10)))^2$.

In Fig. 8 we plot the functional complexity $C_f$ for a 2D mesh versus the bending radius, calculated for different material platforms (different colors). The values used as loss coefficients, $\alpha_s$, $\alpha_t$ and $\alpha_b$ and the length of the tunable element are summarized in Table 1 (the coefficients summarize previous experimental data as described in section 5.4).

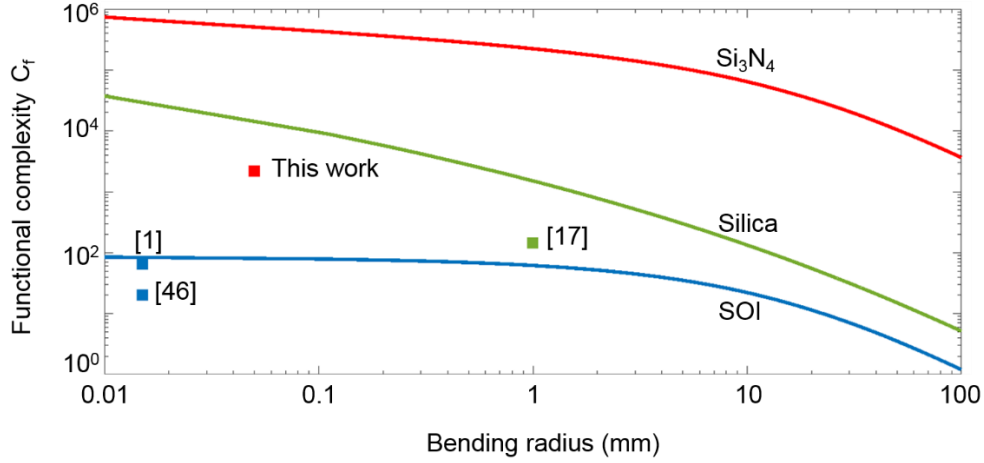

Fig. 8. Functional complexity calculated for three different platforms: silicon nitride (red curve), SOI (blue) and doped silica (green). The highest functional complexity achieved in previous work is indicated as data points, i.e., [1][46] realized in SOI, [17] in doped silica and this work realized with silicon nitride.

It can be seen that silicon nitride provides a functional complexity that is almost four orders of magnitude higher than that of SOI and 1.5 to 3 orders higher than that of silica, depending on the bending radii considered. The much higher propagation loss of SOI is mainly introduced by the tuning via carrier injection [67], even if $L_t$ can be held much shorter than in $Si_3N_4$. With regard to doped silica, the higher functional complexity of silicon nitride is due to its substantially lower bending loss. In order to identify and quantify possible room for improvements we compare three recently published realizations, i.e., [1, 17, 46] and this work, all of them describing on-chip linear optical networks (see data points in Fig. 8). The highest complexity of current SOI processors is close to the maximum possible and that of silica can be improved by up to two orders of magnitude with smaller radii of curvature. The largest room for improvement is expected for $Si_3N_4$. Although the complexity of $Si_3N_4$ is leading already by one order of magnitude, more than three orders of magnitude seem possible.

Table 1. Parameters used for the functional complexity calculation.
*The dependence of bending loss vs. waveguide curvature radius is obtained as described in section 5.4.

|  | Straight-propagation loss $\alpha_s$[dB/m] | Bending loss $\alpha_b$[dB/m] | Phase-tuning loss $\alpha_t$[dB/m] | Tunable elements length $L_t$[mm] |
|---|---|---|---|---|
| SOI | 2.7[69] | $4.07 \cdot R^{-0.62}$ | 3700[67] | 0.0616[67] |
| $Si_3N_4$ | 0.045[27] | $0.316 \cdot R^{-0.95}$ | 0 | 12[70] |
| Doped Silica | 0.01[71] | $7.24 \cdot R^{-0.74}$ | 0 | 5[72] |

*5.2 Photonic processor*

To ensure propagation losses of 0.2 dB/cm, a small footprint and single-mode propagation at telecom C-band wavelengths (around 1550 nm) a double-stripe waveguide cross section is chosen [33]. The total on-chip propagation losses are less than 40%, a value that corresponds to the longest-possible on-chip geometrical path length (about 10 cm, along 15 tunable beam splitters). An array of polarization-maintaining fibers is bonded to the chip to give optical access to the waveguides. The processor is temperature-stabilized by a Peltier element and independent thermal tuning of the 128 phase shifters is accomplished via USB-controlled drivers. The entire assembly comprising the chip, fiber arrays and electronics, is packaged into a single portable case with a USB connection and power socket at its back, and with 16 FC/PC fiber connectors at the front panel. After transportation of the box from Twente to Oxford, there was no need for recalibrating the tuning elements on chip, showing that the assembly is robust against vibrations and insensitive to fluctuations in the environment temperature. After plugging in the single-photon source, the experiments were carried out straightforwardly via computer control of the USB input.

*5.3 High-dimensional single-photon quantum gates*

Figure 9 shows the schematic of how to implement any integer power of a d-dimensional Pauli X-gate with a d-dimensional linear optical network in Reck's scheme [49] using d-dimensional rail encoding. The linear optical network is described by the schematic in Fig. 9 where the element $e_{i,j}$ indicates the tunable beam splitter at the $i^{th}$ row and $j^{th}$ column of the linear optical network. The $n^{th}$ power of a d-dimensional X-gate, i.e., $X^n$, can be found by setting the reflectivity of all the beam splitters $e_{i,j}$ to 1 (bar mode) and changing to 0 (cross mode) the reflectivity of the elements up to the $n^{th}$ column and up to the $((d-n)+(j-1))^{th}$ row, for $j = 1, …,$ n.

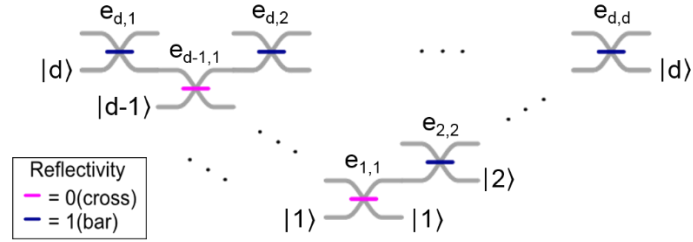

Fig. 9. Schematic of how to implement a d-dimensional X-gate on a d-dimensional linear optical network.

*5.4 Derivation of loss coefficients*

We recall that radiation loss occurs at all waveguide curvatures (bending loss). When keeping the waveguide cross section constant along the propagation coordinate, the bending loss coefficient, $\alpha_b(R)$, increases exponentially with the inverse radius of curvature, R [73]. Using short bending radii reduces the propagation length through bent waveguides, so the bending loss, reducing also the area occupied by a circuit. For bent waveguides with increasing radius of curvature, the loss levels off to the minimum value, $\alpha_s$, given by the straight-propagation loss, which is usually given by material-intrinsic absorption and Rayleigh scattering.

Both the bending loss and straight-propagation loss are strongly dependent on the index contrast between core and cladding as defined by the selected material platform, by the chosen shape and size of the waveguide core. They depend also largely on the chosen fabrication process. Generally, tightly confining the light to the core with high-contrast waveguides and large, wavelength-sized cross sections, reduces the bending loss but simultaneously increases the straight-propagation loss. Weak guiding on the other hand, as achieved with low index contrast or a small (sub-wavelength) core size in high-contrast materials, can yield very low straight-propagation loss; however, the bending loss becomes significant. In conclusion, the overall loss in a circuit is minimized if the waveguide cross section is continuously adjusted to the local curvature radius. This approach can be seen, e.g., in recent work with silicon waveguides [73, 74].

In order to analyze which waveguide platform offers highest component density until the power transmission is reduced to $e^{-1}$ we derive an empirical expression for the minimum value of $\alpha_b(R)$ vs the bending radius.

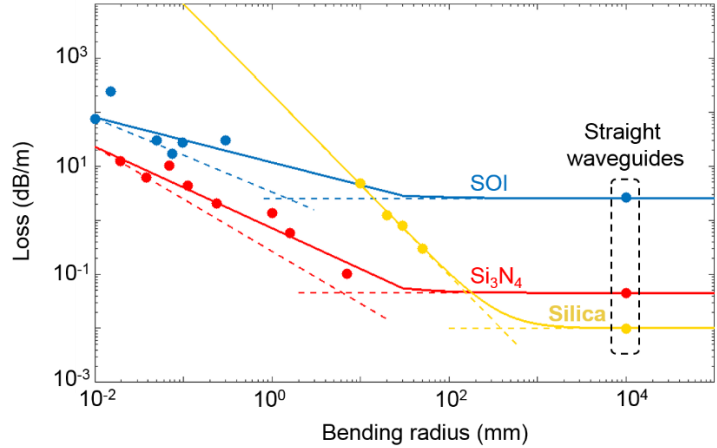

Fig. 10. Overview of planar waveguide propagation loss versus bending radius as in [63] with more recent works [1,61,62,64]. The loss value for large radius tends to $\alpha_s$.

Figure 10 displays experimentally determined loss constants as reported for SOI, $Si_3N_4$ and doped $SiO_2$ (silica) waveguides vs the fabricated bending radius. Data published until 2014 are taken from [75] and more recent data from [74, 76, 77] and [1]. Although the loss values in the named references show a significant variation, selecting only data points with the lowest loss for each radius and platform yields clear trends. Specifically, the minimum reported loss data vs waveguide bending radius shows that the lowest-loss values follow approximately an inverse power law. To be noted is that there might be data points that deviates from the derived power law, e.g., a loss value of ~0.1 dB/m for a bending radius of ~ 115 μm as reported in [28] for $Si_3N_4$. However, these data points are exceptional and do not affect much the slope of the derived power law curves. We note that this dependence was found approximately also in numerical calculations when adjusting the width and height of rectangular waveguides from silicon, $Si_3N_4$, and $GeO_2$-doped silica waveguides [75]. Fitting inverse power laws to the lowest-loss experimental data is thus consistent with the assumption that each lowest-loss experimental observation was based on choosing an optimum waveguide cross section for the corresponding bending radius. The coefficients extracted for each platform from the power-law fits are listed in Table 1.

As reported above, with increasing radius of curvature, the value of $\alpha_b(R)$ levels off to a constant offset value, which is the minimum loss for straight waveguides $\alpha_s$. To indicate these levels, we have drawn in Fig. 10 horizontal dashed and solid lines that pass through the lowest reported experimental loss values (data points in the dashed frame taken from [27] for $Si_3N_4$, from [69] for SOI and from [71] for doped silica). Adding the according loss constants, $\alpha_s$, to the inverse power law functions then yield a closed expression for the expected minimum loss vs bending radius (solid curves in Fig. 10).

Eventually, for a complete description of losses in tunable and programmable photonic processors, also the loss in the phase-tunable sections of a waveguide circuit has to be quantified, $\alpha_t$ (which we term tunability-induced loss). For thermo-optic tuning using large-bandgap dielectric materials, here doped silica and $Si_3N_4$, no additional losses are expected or have been reported. To implement highly effective tuning in semiconductors within short propagation lengths, carrier injection can be applied, however, this increases the propagation loss through free-carrier absorption [67]. For the waveguide cross sections to be used we assume standard values, i.e., 220 nm thickness for SOI waveguides [69], and weakly guiding high-aspect ratio waveguides for $Si_3N_4$ [27] and doped silica [78].

## Funding

## Acknowledgment

We thank Tristan Tentrup for useful discussions.